\pdfoutput=1
\documentclass[11pt,twoside,a4paper,cmspaper,final,collab]{cms-tdr}
\def\svnVersion{2500acd}\def\svnDate{2025/01/27}\def\cmsCernNoTag{CERN-EP-2024-225}\def\cmsCernDate{\today}\def\cmsMessage{Published in Reports on Progress in Physics as \href{http://dx.doi.org/10.1088/1361-6633/adaa13}{\doi{10.1088/1361-6633/adaa13}.}}
\begin{document}\cmsNoteHeader{EXO-23-013}

\newcommand{\iptwo}{\ensuremath{d_{xy}}\xspace}
\newcommand{\ipthr}{\ensuremath{d_{xyz}}\xspace}
\newcommand{\ipsigtwo}{\ensuremath{d_{xy}/\sigma_{xy}}\xspace}
\newcommand{\ipsigthr}{\ensuremath{d_{xyz}/\sigma_{xyz}}\xspace}
\newcommand{\gnnd}{\ensuremath{g_{\text{displaced}}}\xspace}
\newcommand{\gnnp}{\ensuremath{g_{\text{prompt-veto}}}\xspace}
\newcommand{\PG}{\ensuremath{\text{G}_{0}}\xspace}
\newcommand{\mG}{\ensuremath{m_{0}}\xspace}
\newcommand{\mTP}{\ensuremath{m_{\PQT}}\xspace}
\newcommand{\cPS}{\ensuremath{\text{S}}\xspace}
\newcommand{\HSS}{\ensuremath{\PH\to\cPS\cPS}\xspace}
\newcommand{\sbb}{\ensuremath{\cPS\to\bbbar}\xspace}
\newcommand{\sdd}{\ensuremath{\cPS\to\PQd\PAQd}\xspace}
\newcommand{\stt}{\ensuremath{\cPS\to\PGt\PGt}\xspace}
\newcommand{\CHI}{\ensuremath{\chi^{2}/\text{dof}}\xspace}
\newcommand{\mS}{\ensuremath{m_{\cPS}}\xspace}
\newcommand{\ctau}{\ensuremath{c\tau_{0}}\xspace}
\newcommand{\dR}{\ensuremath{\Delta R}\xspace}
\newcommand{\BHSS}{\ensuremath{\mathcal{B}(\HSS)}\xspace}
\newcommand{\BHGG}{\ensuremath{\mathcal{B}(\PH\to\PG\PG)}\xspace}
\newcommand{\abseta}{\ensuremath{\abs{\eta}}\xspace}
\newcommand{\nx}[1]{\ensuremath{N_{\mathrm{#1}}}\xspace}
\newcommand{\nX}{\nx{X}}
\newcommand{\nA}{\nx{A}}
\newcommand{\nB}{\nx{B}}
\newcommand{\nC}{\nx{C}}
\newcommand{\nD}{\nx{D}}
\newcommand{\Rik}{\ensuremath{R_{ik}}\xspace}
\newcommand{\fRtv}{\ensuremath{f_{R_{\text{track-vertex}}}}\xspace}
\newcommand{\xxi}{\ensuremath{x_{i}}\xspace}
\newcommand{\yk}{\ensuremath{y_{k}}\xspace}
\newcommand{\eik}{\ensuremath{e_{ik}}\xspace}
\newcommand{\ykp}{\ensuremath{y_{k}^{\prime}}\xspace}
\newcommand{\fOv}{\ensuremath{f_{O_{\text{vertex}}}}\xspace}

\ifthenelse{\boolean{cms@external}}{\providecommand{\cmsLeft}{upper\xspace}}{\providecommand{\cmsLeft}{left\xspace}}
\ifthenelse{\boolean{cms@external}}{\providecommand{\cmsRight}{lower\xspace}}{\providecommand{\cmsRight}{right\xspace}}
\newlength\figThreeWidth
\ifthenelse{\boolean{cms@external}}{\setlength{\figThreeWidth}{0.41\textwidth}}{\setlength{\figThreeWidth}{0.49\textwidth}}
\ifthenelse{\boolean{cms@external}}{\newcommand{\app}{}}{\newcommand{\app}{Appendix}}

\newlength\cmsTabSkip\setlength{\cmsTabSkip}{1ex}

\cmsNoteHeader{EXO-23-013}
\title{Search for light long-lived particles decaying to displaced jets in proton-proton collisions at \texorpdfstring{$\sqrt{s} = 13.6\TeV$}{sqrt(s) = 13.6 TeV}}
\titlerunning{Search for light long-lived particles decaying to displaced jets at 13.6\TeV}

\date{\today}

\abstract{
A search for light long-lived particles decaying to displaced jets is presented, using a data sample of proton-proton collisions at a center-of-mass energy of 13.6\TeV, corresponding to an integrated luminosity of 34.7\fbinv, collected with the CMS detector at the CERN LHC in 2022. Novel trigger, reconstruction, and machine-learning techniques were developed for and employed in this search. After all selections, the observations are consistent with the background predictions. Limits are presented on the branching fraction of the Higgs boson to long-lived particles that subsequently decay to quark pairs or tau lepton pairs. An improvement by up to a factor of 10 is achieved over previous limits for models with long-lived particle masses smaller than 60\GeV and proper decay lengths smaller than 1\unit{m}. The first constraints are placed on the fraternal twin Higgs and folded supersymmetry models, where the lower bounds on the top quark partner mass reach up to 350\GeV for the fraternal twin Higgs model and 250\GeV for the folded supersymmetry model.
}

\hypersetup{%
pdfauthor={CMS Collaboration},%
pdftitle={Search for light long-lived particles decaying to displaced jets in proton-proton collisions at sqrt(s) = 13.6 TeV},%
pdfsubject={CMS},%
pdfkeywords={CMS, LLP, displaced jets, exotic Higgs decays}}

\maketitle 
\section{Introduction}

Long-lived particles (LLPs) that have macroscopic decay lengths ($\gtrsim$0.1\mm) are ubiquitous in both the standard model (SM) and beyond-the-SM (BSM) scenarios.
Many models of BSM physics naturally predict the production of hadronically decaying LLPs at the CERN LHC, which leads to displaced jets (DJs) whose origins are far from the interaction point.
Examples include, but are not limited to, supersymmetry (SUSY)~\cite{ArkaniHamed:2004fb,Giudice:2004tc,Hewett:2004nw,ArkaniHamed:2004yi,Gambino:2005eh,Arvanitaki:2012ps,ArkaniHamed:2012gw,Fayet:1974pd,Farrar:1978xj,Weinberg:1981wj,Hall:1983id,Barbier:2004ez,GIUDICE1999419,Meade:2008wd,Buican:2008ws,Fan:2011yu,Fan:2012jf}, dark sectors (DSs) or other models with dark matter candidates~\cite{Strassler:2006im,Strassler:2006ri,Han:2007ae,Kaplan:2009ag,Hall:2009bx,Kim:2013ivd,Co:2015pka,Calibbi:2018fqf,Chacko:2005pe,Cai:2008au,Craig:2014aea,Craig:2015pha,Curtin:2015fna,Csaki:2015fba,Alipour-fard:2018mre}, models with heavy neutral leptons~\cite{Atre:2009rg,Drewes:2013gca,Deppisch:2015qwa,Cai:2017mow}, baryogenesis triggered by weakly interacting massive particles~\cite{Cui:2011ab,Cui:2012jh,Cui:2014twa}, and models with a ``high-quality'' axion~\cite{Kamionkowski:1992mf,Hook:2019qoh}.
In such cases, a displaced-jet search is a powerful tool to address numerous long-standing puzzles in particle physics.

In this paper, we present a search for at least one LLP produced in proton-proton ($\Pp\Pp$) collisions at the LHC and decaying inside the inner tracking system of the CMS detector~\cite{CMS:2008xjf,CMS:2023gfb}.
Data used in this search were collected at a center-of-mass energy of 
13.6\TeV in 2022, corresponding to an integrated luminosity of 34.7\fbinv. The target 
signature is a pair of jets, referred to as a dijet, arising from the LLP decay.
Displaced vertices (DVs) can be reconstructed using the displaced tracks associated with the dijet.
The properties of the tracks, DVs, and dijet are used to discriminate between exotic LLP signatures and SM background processes.
We focus on light LLPs with masses $m_{\text{LLP}}\lesssim 60\GeV$ that decay to quarks or tau leptons, as this is an important and largely unconstrained phase space. Heavy LLPs with 
hadronic decays were largely excluded by the previous displaced-jets search~\cite{CMS:2020iwv}, which had significantly lower sensitivity to light LLPs.
This search introduces new trigger, reconstruction, and machine learning techniques to increase sensitivity to light LLPs, bringing up to a factor of 10 improvement compared to previous searches~\cite{CMS:2020iwv,CMS:2021yhb,ATLAS:2021jig,ATLAS:2024qoo}.

The benchmark signature for this search is an exotic decay of the 125\GeV Higgs boson to two long-lived neutral scalars \cPS (\HSS), each of which further decays to a pair of SM fermions.
The Feynman diagram for this process is shown in Fig.~\ref{fig:Higgs_feynman}. We focus on hadronic final 
states, including bottom quarks (\sbb), down quarks (\sdd), and tau leptons ($\cPS\to\PGtp\PGtm$, simply denoted as \stt). 
The \sbb and \sdd decays are chosen as representative of LLP decays to heavy-flavor and light-flavor quarks, and the sensitivity of 
this search is similar for different quark flavor assumptions. 
This benchmark signature is motivated by the Higgs portal scenario, where the Higgs boson acts as a portal to DSs containing new SM gauge singlet particles. 
The Higgs boson, because of its unique status in the SM, often mediates the leading interaction between DSs and the SM sector at LHC energy scales and thus provides a natural portal to DSs~\cite{Patt:2006fw}.
Furthermore, the central role of the Higgs boson in addressing many remaining questions in fundamental physics also suggests that new physics may preferentially couple to it.

A well-motivated version of the Higgs portal scenario is ``neutral naturalness''~\cite{Craig:2014aea,Craig:2015pha,Curtin:2015fna}, realizations of which include the twin Higgs~\cite{Chacko:2005pe}, folded SUSY~\cite{Burdman:2006tz}, and quirky little Higgs~\cite{Cai:2008au} models.
In neutral naturalness, the Higgs boson mass is protected by a global symmetry between the DS and the SM sector, which helps resolve the electroweak hierarchy problem. This scenario also has important implications for the nature of dark matter~\cite{GarciaGarcia:2015fol,Craig:2015xla,Curtin:2021spx,Farina:2015uea,Farina:2016ndq,Prilepina:2016rlq,Hochberg:2018vdo}, dark phase transitions and baryogenesis in the early universe~\cite{Schwaller:2015tja,Barbieri:2016zxn,Fujikura:2018duw,Matsedonskyi:2020kuy,Zu:2023olm}, the origin of neutrino masses~\cite{Batell:2015aha,Csaki:2017spo}, and proposed resolutions for tensions in cosmological measurements~\cite{Prilepina:2016rlq,Bansal:2021dfh}.
The lightest hadronic state of the DS is usually a hidden glueball, which has a suppressed decay back to SM particles through the Higgs portal~\cite{Juknevich:2009ji,Juknevich:2009gg}.
Therefore, the hidden glueball is usually long-lived and decays to DJs, preferring displaced \PQb jets because of mixing between the DS and the SM Higgs boson.
The hidden glueball can be produced in decays of the Higgs boson and therefore equates to the generic \cPS in the benchmark signature.
The hidden glueball is theoretically preferred to have a mass between 10 and 60\GeV~\cite{Craig:2015pha,Curtin:2015fna,deFlorian:2016spz}.
Because of hadronization in the DS, the Higgs boson can decay to more than 
two hidden glueballs, especially when the glueball mass is small~\cite{Batz:2023zef}. Nevertheless, the 
two-body \HSS decay is still the most common decay topology in the glueball 
mass range considered here and is therefore taken as the benchmark.

Although the \HSS decay is chosen as the benchmark,  we do not attempt to reconstruct the Higgs boson candidate or to place specific requirements on the event topology. Instead, we focus on the reconstruction and identification of each LLP. As a result, the search is also sensitive to 
many other BSM scenarios with hadronically decaying LLPs. 

The paper is organized as follows. A brief description of the CMS detector is introduced 
in Section~\ref{sec:CMS}. The simulated samples are described in Section~\ref{sec:sim}. 
Section~\ref{sec:trigger} details the object reconstruction and displaced-jet triggers (DJTs). Section~\ref{sec:DV} describes the DV reconstruction and LLP identification 
algorithms. Section~\ref{sec:sel} describes the event selections and the background estimation method. The systematic uncertainties are presented in Section~\ref{sec:sys}. 
The results and interpretations are described in Section~\ref{sec:res}. The paper is summarized in 
Section~\ref{sec:sum}. Tabulated results are provided in the HEPData record for this analysis~\cite{hepdata}. A glossary of the acronyms used in the paper is provided in \app~\ref{sec:glossary}.

\newlength\figOneWidth
\ifthenelse{\boolean{cms@external}}{\setlength{\figOneWidth}{0.99\linewidth}}{\setlength{\figOneWidth}{0.75\textwidth}}
\begin{figure}[hbtp]
\centering
\includegraphics[width=\figOneWidth]{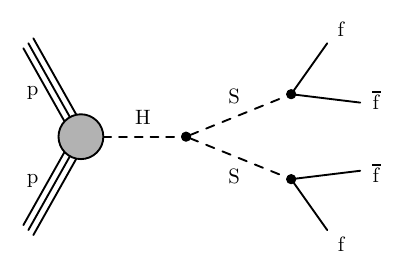}
\caption{The Feynman diagram for the benchmark signal model, in which the SM-like Higgs boson with a mass of $125\GeV$ decays to two long-lived neutral scalars \cPS,
and each of them decays to a pair of SM fermions.}
\label{fig:Higgs_feynman}
\end{figure}

\section{The CMS detector}\label{sec:CMS}

The central feature of the CMS apparatus is a superconducting solenoid of 6\unit{m} internal diameter, providing a magnetic field of 3.8\unit{T}. Within the solenoid volume are a silicon pixel and strip tracker, a lead tungstate crystal electromagnetic calorimeter (ECAL), and a brass and scintillator hadron calorimeter (HCAL), each composed of a barrel and two endcap sections. Forward calorimeters extend the pseudorapidity coverage provided by the barrel and endcap detectors. Muons are measured in gas-ionization detectors embedded in the steel flux-return yoke outside the solenoid.

The silicon tracker measures charged particles within the pseudorapidity range $\abseta < 3.0$. During the LHC running period when the data used in this paper were recorded, the silicon tracker consisted of 1856 silicon pixel and 15\,148 silicon strip detector modules. Details on the pixel detector can be found in Ref.~\cite{Phase1Pixel}. For nonisolated particles with $1 < \pt < 10\GeV$ and $\abseta < 3.0$, the track resolutions are typically 1.5\% in \pt and 20--75\mum in the transverse impact parameter (\iptwo)~\cite{DP-2020-049}.

In the region $\abseta < 1.74$, the HCAL cells have widths of 0.087 in pseudorapidity and 0.087 in azimuth ($\phi$). In the $\eta$-$\phi$ plane, and for $\abseta < 1.48$, the HCAL cells map on to $5{\times}5$ arrays of ECAL crystals to form calorimeter towers projecting radially outwards from close to the nominal interaction point. For $\abseta > 1.74$, the coverage of the towers increases progressively to a maximum of 0.174 in $\Delta \eta$ and $\Delta \phi$. Within each tower, the energy deposits in ECAL and HCAL cells are summed to define the calorimeter tower energies, which are subsequently used to provide the energies and directions of hadronic jets.

Events of interest are selected using a two-tiered trigger system. The first level, composed of custom hardware processors, uses information from the calorimeters and muon detectors to select events at a rate of around 100\unit{kHz} within a fixed latency of about 4\mus~\cite{CMS:2020cmk}. The second level, known as the high-level trigger (HLT), consists of a farm of processors running a version of the full event reconstruction software optimized for fast processing, and reduces the event rate to around 1\unit{kHz} before data storage~\cite{CMS:2016ngn}.

A more detailed description of the CMS detector, together with a definition of the coordinate system used and the relevant kinematic variables, can be found in Ref.~\cite{CMS:2008xjf}. 

\section{Event simulation}\label{sec:sim}

The backgrounds in this search include nuclear interactions, long-lived SM hadrons, and misreconstructed DVs formed by unrelated crossing tracks.
Such phenomena mainly occur in SM events containing jets produced through the strong interaction, referred to as quantum chromodynamics (QCD) multijet events.
The simulated QCD multijet sample is generated at leading order (LO) with \PYTHIA~8.306~\cite{Bierlich:2022pfr}.
Parton showering and hadronization are simulated with \PYTHIA, using the CP5 underlying-event tune~\cite{Sirunyan:2019dfx}. The simulated QCD multijet sample is mainly used to guide the 
analysis strategy, train the LLP-identification taggers, and estimate systematic uncertainties, while 
the background estimation for this search is purely determined from data.

The \POWHEG~2.0~\cite{Nason:2004rx,Frixione:2007vw,Alioli:2010xd,Bagnaschi:2011tu} program is used to generate events containing a 125\GeV Higgs boson via gluon-gluon fusion at 
next-to-LO, which is the dominant production mode of the SM-like Higgs boson.
The process \HSS and the \sbb, \sdd, and \stt decays are then simulated using \PYTHIA.
Signal samples are produced with LLP masses \mS of 15, 23, 30, 40, and 55\GeV, and mean proper decay lengths (\ctau) from 1\mm to 1\unit{m}. The samples are normalized according to the gluon-gluon production 
cross section of a 125\GeV Higgs boson at a center-of-mass energy of 13.6\TeV~\cite{Karlberg:2024zxx}.

Both generators use the NNPDF3.1 next-to-next-to-LO parton distribution functions~\cite{Ball:2017nwa}.
The detailed CMS detector response is modeled with a \GEANTfour-based~\cite{Agostinelli:2002hh} simulation.
The effects of additional $\Pp\Pp$ interactions within the same or nearby bunch crossings (``pileup'') are included.

\section{Object reconstruction and displaced-jet triggers}\label{sec:trigger}

Jets are reconstructed from the energy deposits in the calorimeter towers, clustered using the anti-\kt algorithm~\cite{Cacciari:2008gp,Cacciari:2011ma} with a distance parameter of 0.4.
Identification of the leading primary vertex (PV) is a prerequisite for the selection of DJs.
The leading PV is taken to be the vertex corresponding to the hardest scattering in the event, evaluated using tracking information alone, as described in Section 9.4.1 of Ref.~\cite{CMS-TDR-15-02}.
We consider tracks and jets reconstructed both at the HLT and using the full event reconstruction software, with the former objects used in the DJTs and the latter objects, described as ``offline'', used for analyzing the events collected with the DJTs.

The data were collected with dedicated triggers aimed at selecting events with DJs from LLP decays.
In addition to the DJTs implemented in 2017--2018~\cite{CMS:2020iwv}, new DJTs were also been developed
and implemented in 2022~\cite{CMS-DP-2023-043} to significantly improve the trigger efficiencies for light LLPs.
Displaced jets are identified, or ``tagged'', in the HLT using two different requirements.
The first, referred to as the ``inclusive'' tagging requirement, requires that the jet has at most one associated prompt track.
Prompt tracks are defined to have $\pt>1\GeV$, a \iptwo with respect to the
leading PV smaller than 0.5\mm, and a \iptwo significance (\ipsigtwo, the ratio of \iptwo and its uncertainty) smaller than 5.0.
The second, referred to as the ``displaced'' tagging requirement,
encompasses the inclusive tagging requirement and additionally requires that if there is exactly one associated prompt track, there should also
be at least one associated track with $\pt>1\GeV$ and $\iptwo>0.3\mm$.

Two DJTs are implemented.
The first trigger requires $\HT>430\GeV$, where \HT is the scalar \pt sum of all jets with $\pt>40\GeV$ and $\abseta<2.5$.
The trigger also requires the presence of at least two jets satisfying $\pt>40\GeV$, $\abseta<2.0$, and the inclusive tagging requirement.
The events selected by this trigger are further required to have an offline $\HT>450\GeV$ to make sure the online \HT requirement reaches full efficiency and the difference between the observed and simulated efficiencies is negligible.
The second trigger is seeded by a first-level trigger that requires $\HT>240\GeV$ and the presence of a muon with $\pt>6\GeV$, in order to improve the trigger efficiencies for LLPs with heavy-flavor decays.
The trigger further requires that there are at least two jets satisfying $\pt>40\GeV$, $\abseta<2.0$, and the displaced tagging requirement.
The events selected by this trigger are further required to have an offline $\HT>240\GeV$.

The overall trigger efficiencies are $\approx$0.4--1.0\% for \sbb and \sdd in most of the considered 
mass and lifetime range. These efficiencies degrade to $\approx$0.2--0.7\% for \stt, because of the nonzero branching fraction for tau leptons to decay leptonically. Compared to the DJTs implemented in 2017--2018, the new trigger efficiencies are a factor of 4--17 higher for the \HSS, \sbb signature in the parameter ranges explored here.

\section{Displaced-vertex reconstruction and long-lived particle identification}\label{sec:DV}

After the trigger selections, dijet candidates are formed from all possible pairs of offline jets in the event using the jets with $\pt>40\GeV$ and $\abseta<2.0$.
The track candidates used in this search are required to satisfy $\pt>1\GeV$ and pass the high-purity selection~\cite{Chatrchyan:2014fea}.
For a given dijet candidate, the track candidates associated with each jet are selected by requiring that $\dR<0.5$, where $\dR = \sqrt{\smash[b]{(\Delta\eta)^2+(\Delta\phi)^2}}$ is the angular distance between the jet axis and the track direction.
When a track satisfies $\dR<0.5$ for both jets in the dijet candidate, it is associated with the jet with the smaller \dR.

For each dijet candidate, DVs are reconstructed using the associated displaced tracks that satisfy $\iptwo>0.5\mm$ and $\ipsigtwo>5.0$.
Two DV reconstruction approaches are taken.
The first approach is to directly apply adaptive vertex reconstruction (AVR)~\cite{Waltenberger:2008zz,Strandlie:2010zz}, which is an iterative application of the adaptive vertex fitter (AVF)~\cite{Fruhwirth:2007hz}, to all the associated displaced tracks. This approach efficiently reconstructs the LLP decay vertex. 
The second approach is to cluster the displaced tracks first, based on their distances of closest approach and points of closest approach (PCAs), which improves the efficiency to 
reconstruct the additional DVs arising from processes such as \cPqb hadron decays in \sbb.
During the clustering, each displaced track is treated as a seed track, and a cluster is then formed by examining the compatibility between the seed track and other displaced tracks based on the distances of closest approach, PCAs, and the
angles between the track directions and the PV-to-PCA direction.
The AVR is then applied to each cluster of tracks, resulting in another set of DVs.
The DVs from the two approaches are then combined to form a single set of DVs.
During the initial reconstruction, some displaced tracks originating from a given DV 
may not be associated with it after the vertex fitting. 
To mitigate this effect, all the displaced tracks and DVs are reexamined, and each DV is refitted with the AVF using all the displaced tracks that have a three-dimensional impact parameter significance ($\ipsigthr$, the ratio of the three-dimensional impact parameter \ipthr and its uncertainty) smaller than 5.0 with respect to the DV.
After the refitting, only the vertices with a $\chi^{2}$ per degree of freedom (\CHI) smaller than 5.0 are kept.
Since there can be overlaps in the results of the two DV reconstruction approaches, there may be some duplicated vertices after the refitting. To account for this, a given DV is removed from the final DV list if it shares at least $20\%$ of its tracks with another DV and the significance of the distance between the two DVs is smaller than 3.0.

With the dijet candidates and their associated tracks and DVs as inputs, we employ graph neural networks (GNNs)~\cite{10.5555/3157382.3157601,pmlr-v70-gilmer17a} as taggers to identify the dijets arising from LLP decays.
The tracks and DVs associated with a given dijet candidate can naturally form graphs, where the 
tracks and DVs can be viewed as nodes of the graphs, while the track-to-vertex, 
vertex-to-track, and track-to-track 
relations can be viewed as the edges that connect two nodes. 
For each type of relation, a relation function can be built using the 
node and edge features. For example, for track-to-vertex relations \Rik, 
they can be described by a relation function \fRtv:
\begin{equation}
\Rik \equiv \fRtv \left(\xxi, \yk, \eik\right),  
\end{equation}
where \xxi represents track features, \yk represents DV features, \eik represents track-to-vertex edge features like the track-to-DV association, and $i$ ($k$) 
is the index of the tracks (DVs). 
The DV features can then be updated based on the track-to-vertex relations using the message passing formalism~\cite{pmlr-v70-gilmer17a}:
\begin{equation}
\ykp = \fOv \left(\yk,\ \sum_{i} \Rik\right), 
\end{equation} 
where \ykp represents the updated DV features, and \fOv is referred to as the vertex objective function. The functions \fRtv and \fOv are learned during the GNN training. 
The track-to-track and vertex-to-track relations are similarly incorporated and applied to track features. 
The updated DV and track features are further processed to provide the discrimination between the LLP signature and background processes. 

Two GNN-based displaced-dijet taggers are implemented.
The first one, referred to as the ``displaced'' tagger, only takes as input the associated displaced tracks and DVs.
The second one, referred to as the ``prompt-veto'' tagger, only takes as input the associated tracks with $\iptwo<0.3\mm$.
The two taggers cover complementary characteristics of exotic LLPs: the presence of DVs from the LLP decays and the lack of prompt particles accompanying the LLP production.
Moreover, the two taggers have negligible correlations for SM background processes, which is verified 
using simulated QCD multijet events, and thus they enable the estimation of the background yield from data.

For the displaced tagger, the input displaced tracks and input DVs are sorted according to their \ipsigtwo values in descending order.
The input displaced track features include \ipsigtwo; \ipsigthr; the distance from the PV to the crossing point of the track helix and
the dijet direction in the transverse plane; whether the track is associated with the jet with a 
larger \pt; and the ratio between the track energy and the energy sum of the tracks associated with the dijet.
The input DV features include the vertex invariant mass and \pt; the vertex track multiplicity; the transverse decay length significance; \CHI of the vertex fit; and the angles between the
vertex momentum vector, the direction from the PV to DV, and the dijet candidate momentum direction.
To characterize the track-to-vertex relations, the track-to-vertex edge features of the input graph for each track-DV pair are taken to be the track-to-DV association; \ipthr and \ipsigthr between the track and the DV; and the angle between the track direction and the direction of the DV displacement from the PV.
The displaced tagger is built with two successive GNN blocks that update the track and DV features with track-to-track, track-to-vertex, and vertex-to-track relations, using the message passing formalism~\cite{pmlr-v70-gilmer17a}.
The outputs of this step are further processed with two parallel long short-term memory (LSTM) layers~\cite{Hochreiter:1997yld}, which treat the 
processed track features and DV features as sequential data, representing decay chains within the LLP decay system. 
The LSTM outputs are further concatenated with two dijet global features: the total number of the DVs and the sum of \ipsigtwo for the tracks associated with DVs.
The concatenated dijet, displaced track, and DV features are processed with a fully connected dense network to produce the final prediction score \gnnd.

For the prompt-veto tagger, the input tracks are sorted according to \ipsigtwo in ascending order. The input track features include \iptwo; \ipsigtwo; the track-to-PV associations for the leading PV and pileup vertices;
the track-to-jet associations; and the ratio between the track energy and the total 
energy of the dijet candidate.
In addition, for each pair of tracks, whether the two tracks are associated with the same PV is taken as the track-to-track edge feature for the input graph.
The input track features are updated with a GNN message passing block that acts on the track-to-track relations.
The outputs of the GNN block are processed with a single LSTM layer and then further processed with a fully-connected dense network, producing the final prediction score \gnnp.

\begin{figure}[hbtp]
\centering
\includegraphics[width=0.49\textwidth]{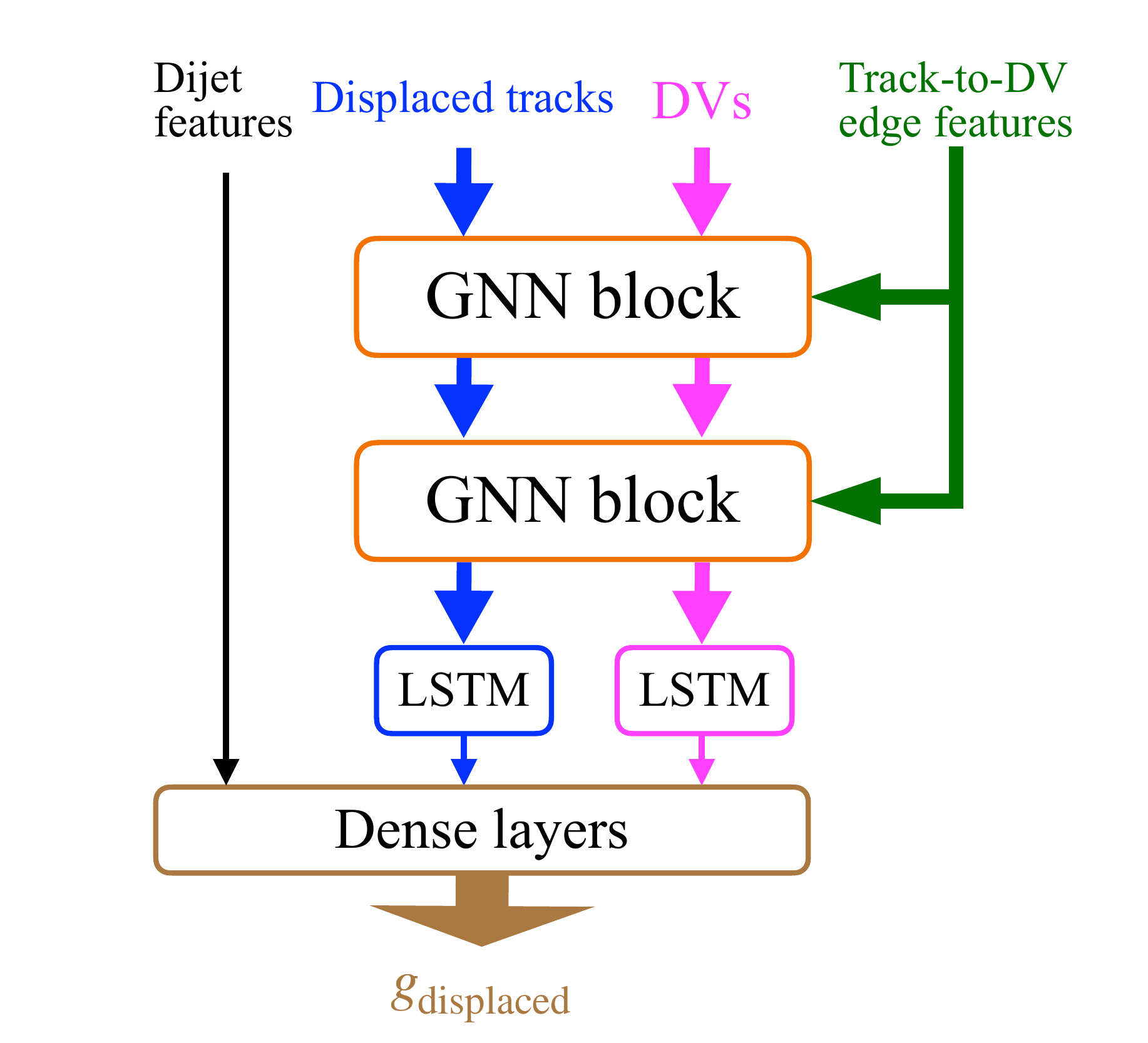}
\includegraphics[width=0.49\textwidth]{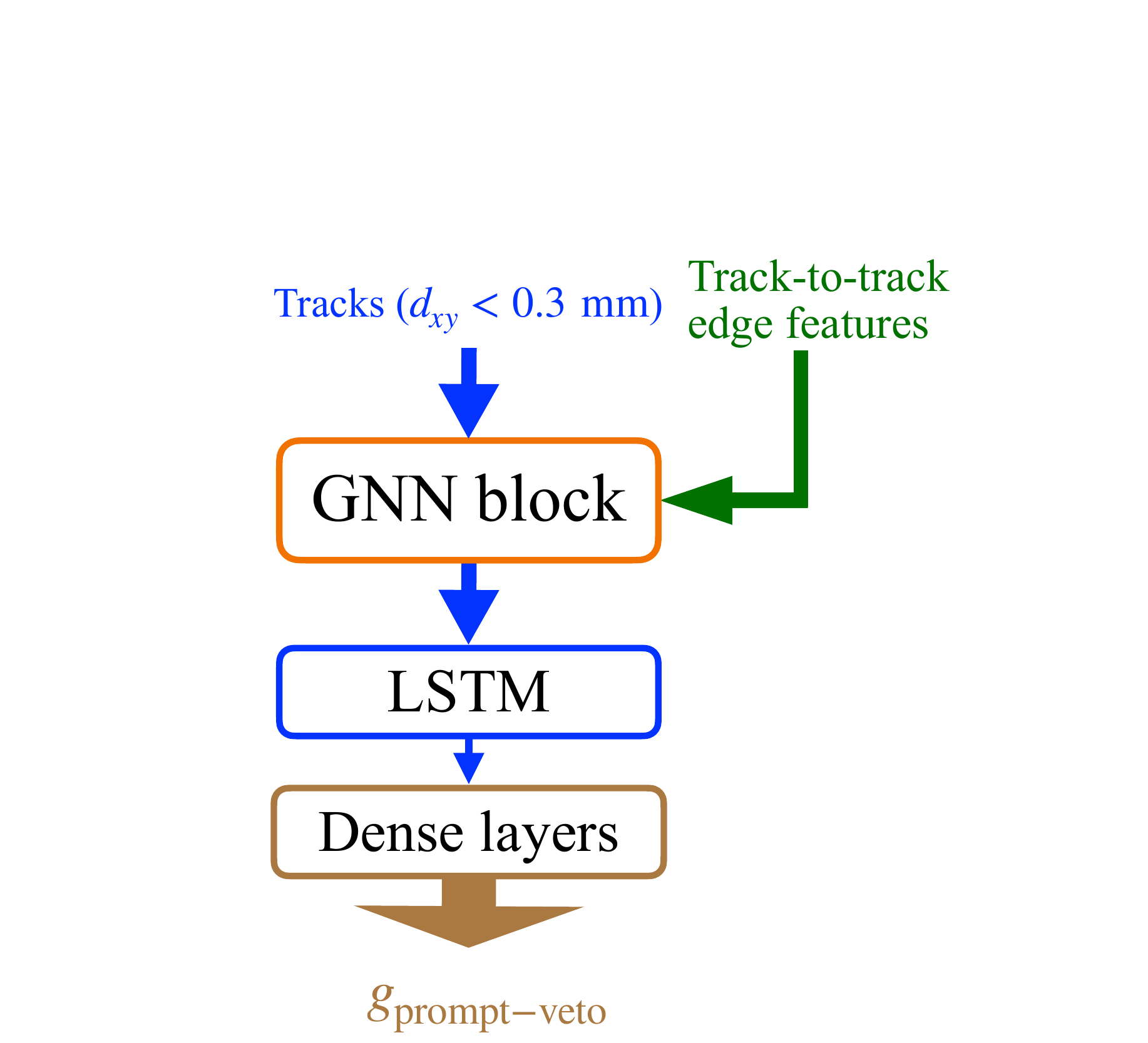}
\caption{The architectures of the displaced (\cmsLeft) and prompt-veto (\cmsRight) taggers. The displaced tagger takes as input the dijet global features, 
displaced tracks, and DVs. The prompt-veto tagger takes as input the tracks with $\iptwo<0.3\mm$.}
\label{fig:GNN_arch}
\end{figure}

The architectures of the displaced and prompt-veto taggers are illustrated in Fig.~\ref{fig:GNN_arch}. Both taggers are implemented and trained using the \textsc{TensorFlow}~v2.6.0 package~\cite{10.5555/3026877.3026899}.
For the training, the binary cross entropy~\cite{Good:1952} is used as the loss function, which is defined as:
\ifthenelse{\boolean{cms@external}}
{
  \begin{multline}
    \mathrm{BCE} =\\
    -\frac{1}{N}\sum_{j=1}^{N}\left[y_{j}\log\left(f(x_{j})\right) + (1-y_{j})\log\left(1-f(x_{j})\right)\right],
    \end{multline}
}
{
  \begin{equation}
  \mathrm{BCE} = -\frac{1}{N}\sum_{j=1}^{N}\left[y_{j}\log\left(f(x_{j})\right) + (1-y_{j})\log\left(1-f(x_{j})\right)\right],
  \end{equation}

}
where $x_{j}$ represents an input, $y_{j}$ represents the class defined as 0 if the input 
is from the background sample or 1 if the input is from 
the signal sample, and $f(x_{j})$ is the prediction of the 
GNN tagger. The simulated QCD multijet sample is used as the background sample.
Simulated \HSS samples with \sbb, $\mS=30$, 40, and 55\GeV, and $\ctau=1$, 10, 100, and $1000\mm$ are combined to form the signal sample.
In a given signal event, the dijet candidate that is the most compatible with the generated LLP decay is selected for the training, according to the
distance between the reconstructed DVs and the generated LLP decay vertex, as well as the angular distance between the momenta of the dijet candidate and the generated LLP.
The GNNs are trained for the \sbb signature because this is the most important decay channel in the Higgs portal scenarios, and the resulting GNNs are also directly applied to the signatures with other decay channels. 
The DJT selections are applied to the events used for the training, and both jets in each dijet candidate are required to satisfy $\pt>40\GeV$ and $\abseta<2.0$. Event weights are assigned during the training such that the total 
sums of the weights are identical for the signal and background samples. The output scores of the GNN taggers range between 0 and 1, with larger values indicating that the dijet is more likely to arise from an LLP decay. 

In general, the displaced tagger achieves a background rejection factor of $10^{4}$ when the signal efficiency is ${\approx}55\%$, while the prompt-veto tagger can achieve a background rejection factor of $10^{3}$ when the signal efficiency is ${\approx}30\%$.
The agreement between data and simulation is verified for the input variables and the GNN output scores, using the events collected with a prescaled control trigger that requires $\HT>425\GeV$.

\section{Event selection and background estimation}\label{sec:sel}

In this search, after the trigger selections, we select the dijet candidates that have at least one reconstructed DV with $\CHI<5.0$.
The displaced and prompt-veto tagger scores are computed for each dijet candidate, and the one with the largest \gnnp in a given event is selected.
We then require that the selected dijet candidate satisfies $\gnnd>0.9985$ and $\gnnp>0.985$, determined by maximizing the 5-standard-deviation discovery potential for the \HSS, \sbb signature using
the Punzi formula~\cite{Punzi:2003bu}, according to the expected signal efficiencies and background yields after the selection.

We define four exclusive regions to employ the ``ABCD'' background estimation method~\cite{Choi:2019mip}.
\begin{itemize}
\item Region A: events with $0.95<\gnnd < 0.9985$, $0.95<\gnnp< 0.985$;
\item Region B: events with $0.95<\gnnd < 0.9985$, $0.985<\gnnp < 1.0$;
\item Region C: events with $0.9985<\gnnd < 1.0$, $0.95<\gnnp < 0.985$; and
\item Region D, the signal region (SR): events with $0.9985<\gnnd < 1.0$, $0.985<\gnnp < 1.0$.
\end{itemize}
The estimated background yield in the SR is thus:
\begin{equation}
\nD^{\text{exp}}=\nB\nC/\nA,
\label{eq:ABCD}
\end{equation}
where \nX is the event yield in region X. In Eq.~\eqref{eq:ABCD}, it is assumed that there is no signal contribution to regions A--C. When interpreting the results for a given signal point, signal 
contributions are taken into account by performing a simultaneous fit in all regions A--D, which will be 
discussed in Section~\ref{sec:results}.

The predicted background yields and the number of observed events in the SR are shown in Fig.~\ref{fig:back_sig}. The
uncertainties in the predicted background yields come from the statistical uncertainties in regions A, B, and C.
Predictions and observations are also shown for regions with smaller \gnnd ranges below 0.9985, achieved by corresponding adjustments to the \gnnd boundaries of regions A--D, in order to validate the background estimation method. The predictions and observations are summarized in Table~\ref{tab:back}. For each observation, a $p$-value is computed based on the lower tail of a Poisson 
distribution convolved with a normalized Gaussian function for the statistical uncertainty of the background prediction. The $p$-value is then converted to a $Z$-value according to the Gaussian error 
function, which represents the observed significance with an equivalent number of standard deviations~\cite{pvalue}. 
The $Z$-values are also listed in Table~\ref{tab:back}, with absolute values all below 1.3 standard deviations, indicating good agreement between the predictions and observations.

\begin{table*}[htb]
\centering
\topcaption{The predicted background yields and observations in the region with $\gnnp>0.985$ for different \gnnd ranges. The background predictions are shown with their statistical uncertainties. The significance of any deviation between the observation and prediction for each \gnnd range is shown as a $Z$-value.}
\label{tab:back}
\begin{tabular}{lccc}
\gnnd  & Predicted background & Observation & $Z$-value\\\hline
(0.96, 0.97) & $\,\,\,68.39\pm12.60$ & 52 & $-1.06$\,\,\\
(0.97, 0.98) & $67.55\pm 9.46$ & 77 & 0.80\\
(0.98, 0.99) & $76.18\pm 8.95$ & 72 & $-0.27$\,\,\\
(0.99, 0.995) & $38.82\pm 5.08$ & 45 & 0.84\\
(0.995, 0.998) & $25.41\pm 3.87$ & 26 & 0.22\\
(0.998, 0.9985) & $\,\,\, 2.83\pm 1.17$ & \, 5 & 1.25\\
(0.9985, 1.0) & $\,\, \,3.34\pm1.28$ & \, 3 & 0.19\\ 
\end{tabular}
\end{table*}

The background estimation was additionally validated using simulated QCD and signal events, both with and without the effect of signal contamination, as well as using observed data in signal-depleted regions with small \gnnp values.
In all cases, the estimated background and the observed value agree within statistical uncertainties.
Therefore, no additional systematic uncertainty is assigned to the predicted background yields. 

\newlength\figTwoWidth
\ifthenelse{\boolean{cms@external}}{\setlength{\figTwoWidth}{0.99\linewidth}}{\setlength{\figTwoWidth}{0.75\textwidth}}
\begin{figure}[htb]
\centering
\includegraphics[width=\figTwoWidth]{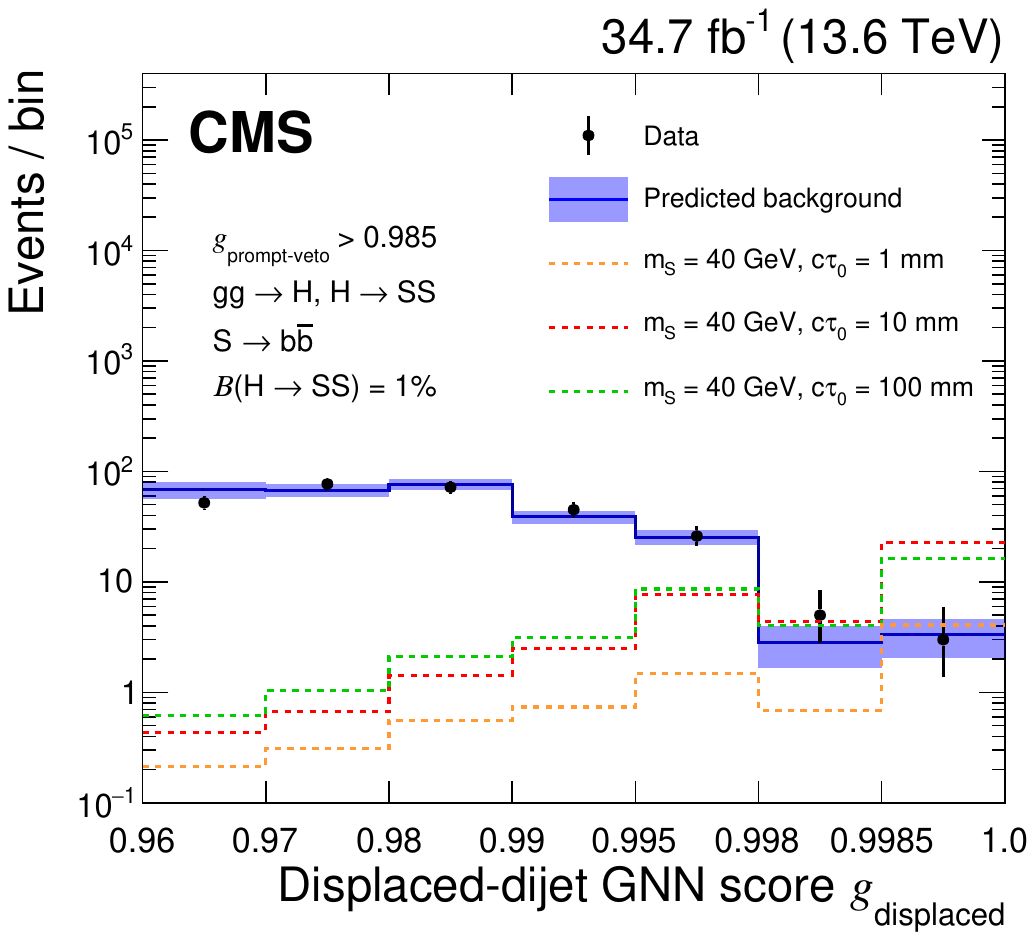}
\caption{The predicted background yields and the number of observed events for the data with $\gnnp>0.985$, shown for different bins of the displaced-dijet GNN score \gnnd. Expected signal yields for the \HSS, \sbb signature are also shown for models with $\mS=40\GeV$ and $\ctau=1$, 10, or 100\mm, assuming a branching fraction of 1\% for the \HSS decay.}
\label{fig:back_sig}
\end{figure}

\section{Systematic uncertainties}\label{sec:sys}

The systematic uncertainty in the integrated luminosity for 13.6\TeV $\Pp\Pp$ collision data in 2022 
is 1.4\%~\cite{CMS:2024sxz}, which is taken as one of the systematic uncertainties in the signal yield.
The systematic uncertainty arising from the pileup modeling is estimated by varying the inelastic $\Pp\Pp$ cross section by 4.6\%~\cite{Sirunyan:2018nqx};
the resulting variation in the signal yield is found to be 1--8\% and is taken as the corresponding systematic uncertainty.
The jet energy scale uncertainty is propagated to the simulated signal samples by varying the jet energy and \pt by one standard deviation,
and the resulting systematic uncertainty in the signal yield is found to be 5--10\%.

Measurements of the efficiency of the online requirements of the DJTs are compared between data and the simulated QCD multijet sample.
Events collected with an isolated single-muon trigger are used for the online \HT requirement, and events collected with a prescaled \HT trigger requiring $\HT>425\GeV$ are used for the online jet \pt requirement.
The differences in the measured efficiencies 
between data and simulation are found to be small, and their impacts on the predicted signal yields are negligible,
so no corresponding uncertainty is assigned.

To estimate the systematic uncertainty in the predicted signal yields from the online tracking requirement of the DJTs, the per-jet efficiencies 
of this requirement are measured as functions of the number of offline prompt tracks and displaced tracks, using events collected with the prescaled \HT trigger.
The difference between data and simulated QCD multijet events is applied to the simulated signal events as a bias for the probability of a single jet to pass the online tracking requirements.
The variation of the efficiency for signal events to have at least two jets passing the online tracking requirements is found to be 0.1--5\% and is taken as the corresponding systematic uncertainty.

The impact of the possible mismodeling of the GNN scores in the simulation on the predicted signal yields is measured similarly, using events collected with the prescaled \HT trigger.
The GNN scores in the simulated samples are varied by the magnitude of the measured discrepancy between data and QCD multijet simulation.
The effect on the signal yields is found to be 4--14\% and is taken as the corresponding systematic uncertainty.

The various systematic uncertainties in the signal yields are summarized in Table~\ref{tab:sys_unc}.

\begin{table*}[htb]
\centering
\topcaption{Summary of the systematic uncertainties in the signal yields.}
\label{tab:sys_unc}
\begin{tabular}{lc}
Source & Uncertainty (\%) \\\hline
Integrated luminosity & 1.4 \\
Pileup modeling & 1--8\\
Jet energy scale & 5--10 \\
Online tracking requirements & 0.1--5\\
GNN modeling & 4--14 \\
\end{tabular}
\end{table*}

\section{Results}\label{sec:res}
The observed event yields in regions A, B, and C are $\nA=722$, $\nB=344$, and $\nC=7$, respectively. The predicted background yield in the SR with $\gnnd>0.9985$ and $\gnnp>0.985$ is therefore $3.34\pm1.28$. We observe 3 events, which is consistent with the prediction. The signal efficiencies 
for representative \HSS signal points in the SR can be found in Table~\ref{tab:eff_HSS}.

\begin{table*}[hbt]
  \topcaption{Signal efficiencies scaled by a factor of $10^{4}$ for the \HSS signature with \sbb, \sdd, and \stt decays in the signal region D, shown for representative signal points with different
\mS and \ctau values. Only statistical uncertainties are listed.}
\label{tab:eff_HSS}
\centering

\begin{tabular}{lccccc}
\multirow{2}*{Decay channel} & \multirow{2}*{\mS (\GeVns)} & \multicolumn{4}{c}{\ctau}\\
&           &      1\mm & 10\mm & 100\mm & 1000\mm \\ \hline

\multirow{3}*{\sbb} &55          & $2.82\pm 0.06$ & $15.46\pm 0.14$ & $12.52\pm 0.12$ & $2.17\pm 0.05$\\
& 40                             & $2.25\pm 0.05$  & $11.96\pm 0.12$ & $8.60\pm 0.10$ & $1.06 \pm 0.04$\\
& 23          & $0.48\pm0.02$ & $4.42\pm 0.07$ & $2.71\pm 0.06$ & $0.20\pm0.01$\\ [\cmsTabSkip]

\multirow{3}*{\sdd} & 55 & $2.80\pm0.06$ & $12.48\pm0.13$ & $10.30\pm0.11$ & $1.89\pm0.05$\\
                    & 40 & $2.47\pm0.05$ & $11.76\pm0.12$ & $8.13\pm0.09$ & $1.06\pm0.04$\\
                    & 23 & $0.59\pm0.03$ & $5.14\pm0.07$ & $2.89\pm0.06$ & $0.19\pm0.02$\\ [\cmsTabSkip]

\multirow{3}*{\stt} & 55 & $0.28\pm0.02$ & $2.17\pm0.05$ & $1.40\pm0.04$ & $0.24\pm0.02$ \\
                    & 40 & $0.23\pm0.02$ & $1.77\pm0.05$ & $1.08\pm0.04$ & $0.15\pm0.01$ \\
                    & 23 & $0.09\pm0.01$ & $0.75\pm0.03$ & $0.39\pm0.02$ & $0.023\pm0.005$\\
\end{tabular}
\end{table*}

\subsection{Interpretations of the results}\label{sec:results}

Upper limits at 95\% confidence level (\CL) are set on the branching fraction \BHSS for different signal models, computed using the \CLs criterion~\cite{Junk:1999kv,Read:2002hq}, with an LHC-style profile likelihood ratio~\cite{CMS-NOTE-2011-005} as the test statistic.
Systematic uncertainties are incorporated through the use of nuisance parameters, 
which are profiled according to the frequentist paradigm. 
The asymptotic approximation~\cite{Cowan:2010js} is used for calculating the \CLs values and has been verified with full-frequentist results for 
representative signal points. 
To account for possible signal contributions in regions A--C, a simultaneous fit is performed for the signal strength and the background yields in all regions A--D, enforcing the ABCD relationship from Eq.~\eqref{eq:ABCD} for the background yields, 
while allowing for a signal component in all regions with relative proportions 
dictated by the signal simulation. The differences between the results of the simultaneous fit and those obtained only using the yields in the SR are smaller than 10\%. These calculations are performed using the \textsc{Combine} package~\cite{CMS:2024onh}.

\begin{figure*}[hbtp]
\centering
\includegraphics[width=0.49\textwidth]{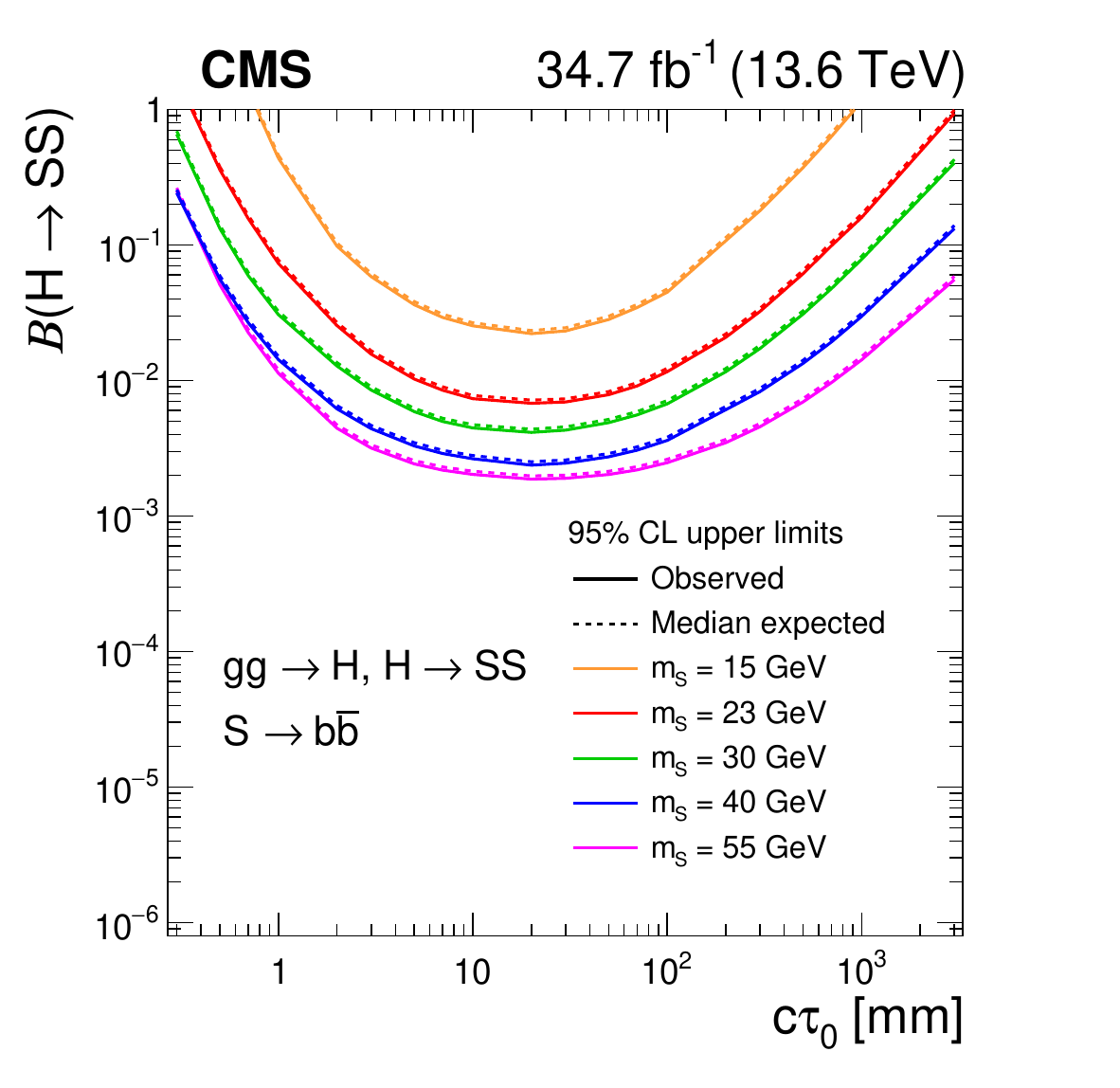}
\includegraphics[width=0.49\textwidth]{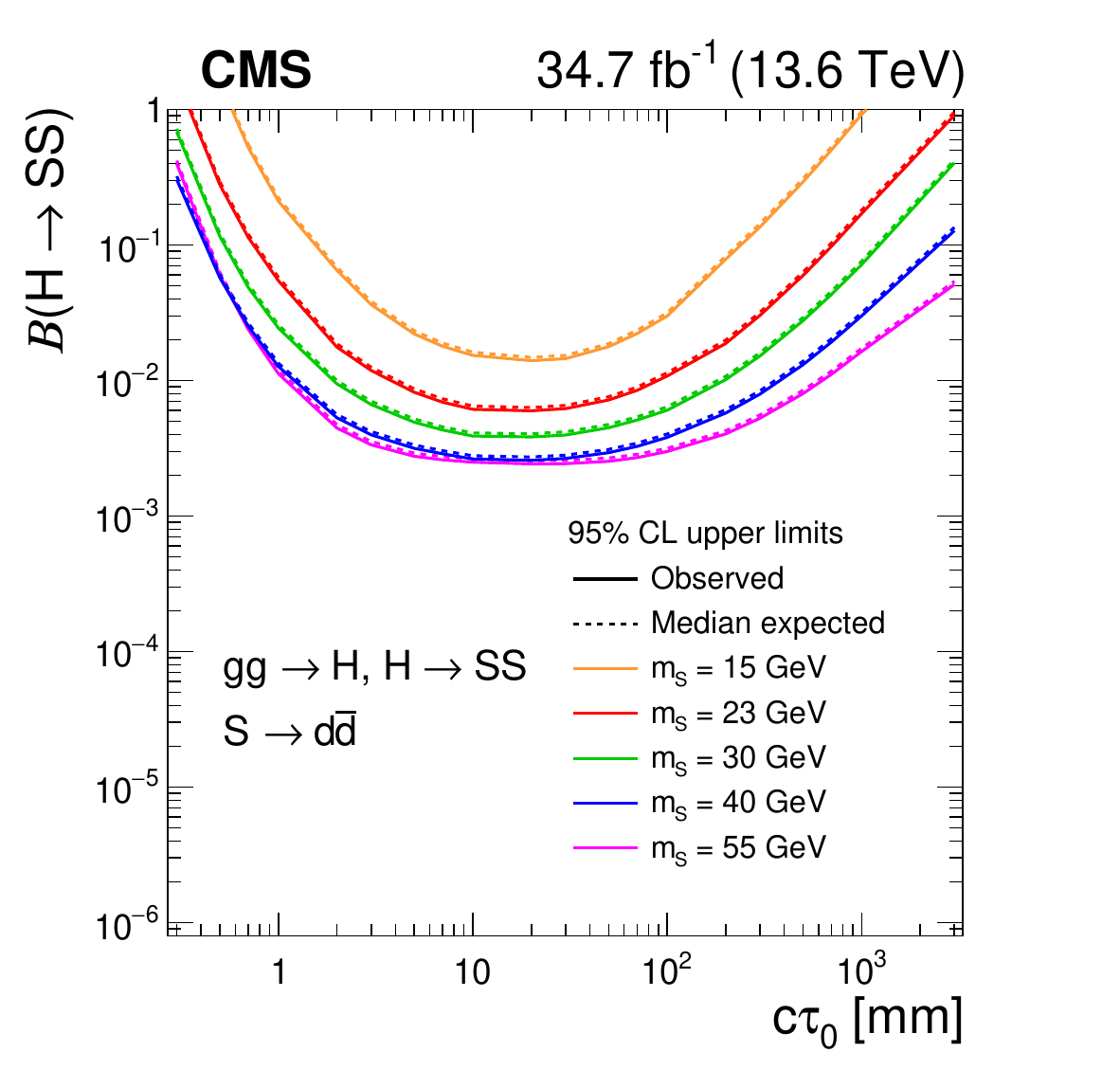}\\
\includegraphics[width=0.49\textwidth]{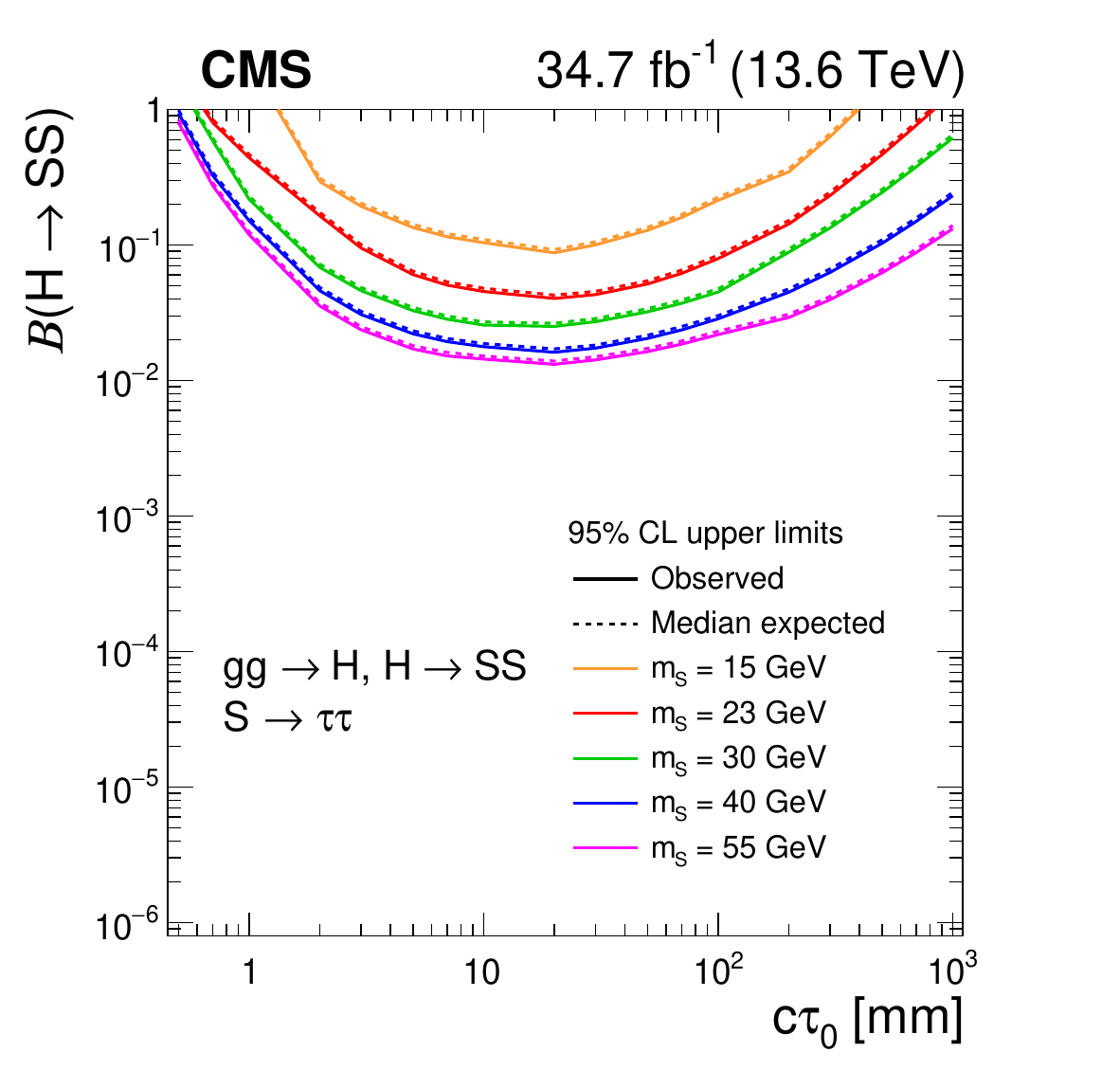}
\caption{The 95\% \CL upper limits on the branching fraction \BHSS for \sbb (upper left), \sdd (upper right), and \stt (lower), for different LLP masses \mS and proper decay lengths \ctau. The 
solid (dashed) lines represent the observed (median expected) limits.}
\label{fig:limits}
\end{figure*}

The upper limits on \BHSS for the \sbb, \sdd, and \stt decay scenarios are shown in Fig.~\ref{fig:limits} for different \mS and \ctau values. 
The limits become weaker for smaller \ctau because only displaced tracks are used to reconstruct DVs,
and for larger \ctau because the tracking efficiency decreases with increasing displacement of the displaced tracks.
The limits also become less stringent at smaller \mS because the boost of the LLP decay system increases, so the two quarks or hadronically decaying tau leptons are more likely to be reconstructed as a single jet. 

The upper limits for the \sbb decay scenario are within 20\% of the \sdd limits when $\mS>20\GeV$, while in the previous search~\cite{CMS:2020iwv} the 
\sbb limits were much weaker than the \sdd limits. The difference in sensitivity between the \sbb and \sdd final states is mitigated by the new DV reconstruction algorithm
and the GNN taggers, which better capture information from the \PQB hadron decay vertices. The new DJTs accept more signal events,
increasing the signal yield compared to previous searches even in this smaller data set.
The upper limits for the \stt decay scenario are weaker than the \sbb and \sdd limits because there is less displaced activity in \stt decays, as there is no hadronization when tau leptons are produced at 
the LLP decay vertex.

\begin{figure*}[hbtp]
\centering
\includegraphics[width=0.49\textwidth]{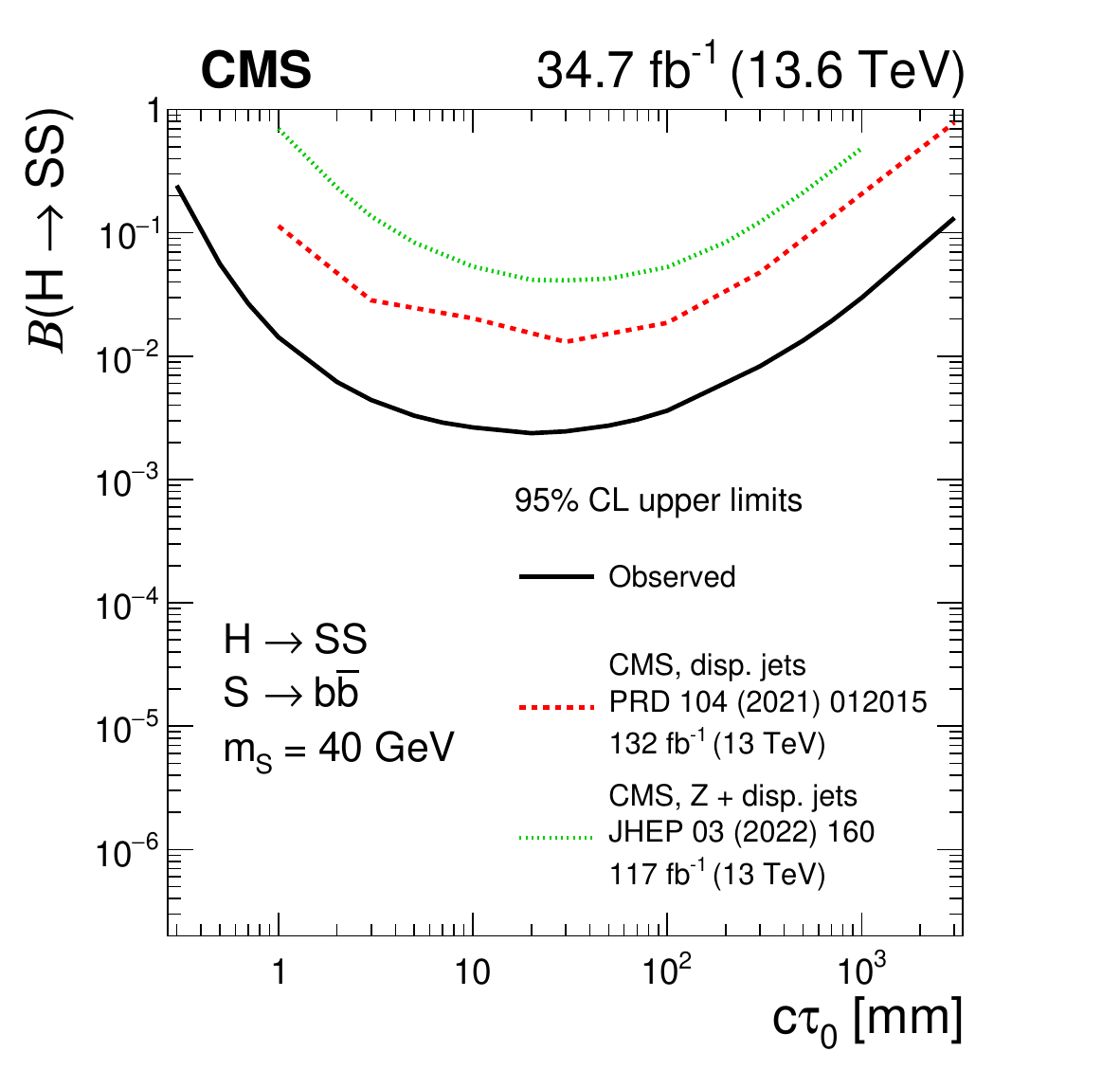}
\includegraphics[width=0.49\textwidth]{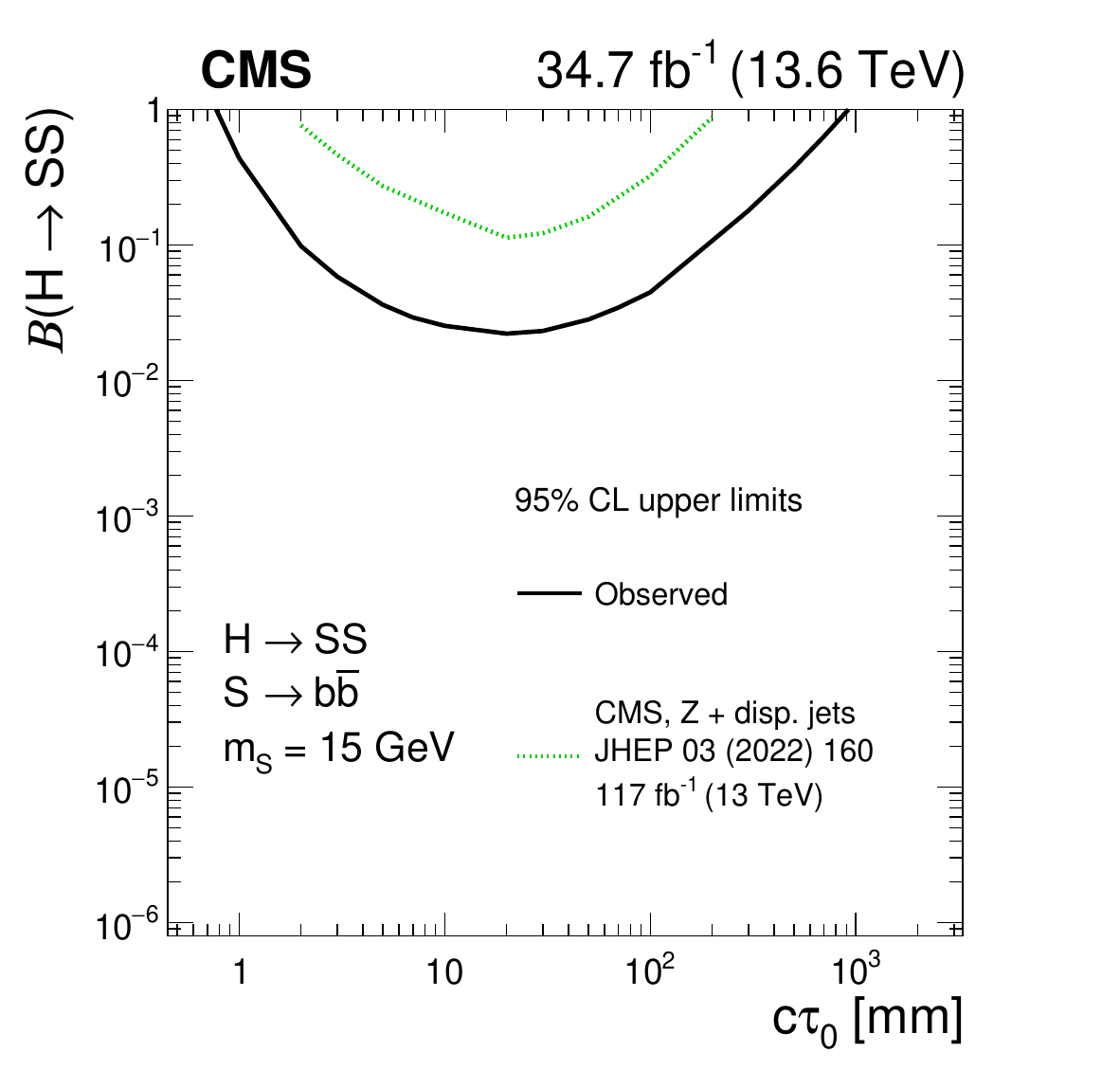}\\
\includegraphics[width=0.49\textwidth]{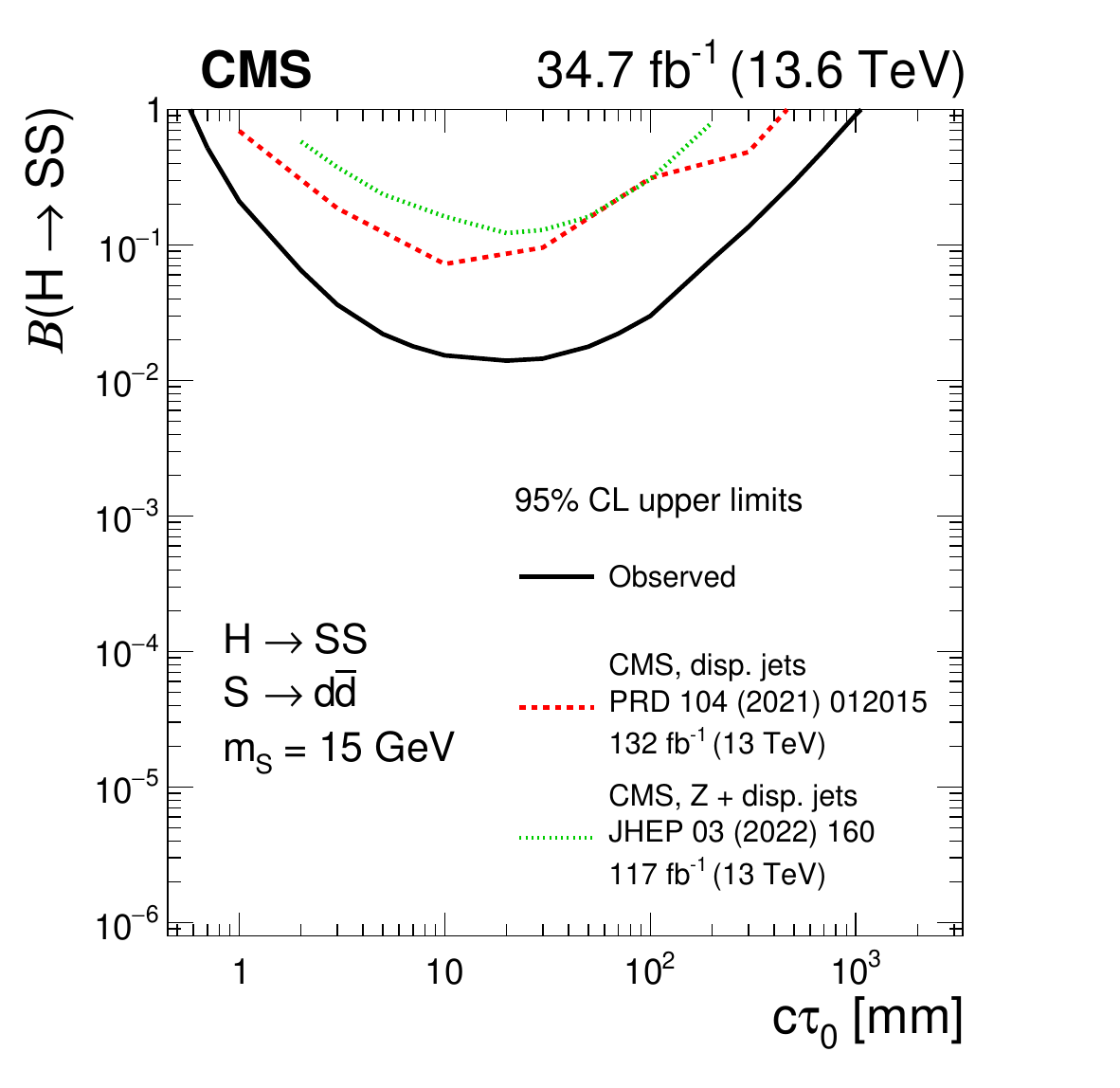}
\caption{Comparisons of the observed limits from this search and other results, for \sbb, $\mS=40\GeV$ (upper left); \sbb, $\mS=15\GeV$ (upper right); and \sdd, $\mS=15\GeV$ (lower). The other results include the previous CMS displaced-jets search~\cite{CMS:2020iwv} (red dashed lines) and the CMS \PZ + displaced-jets search~\cite{CMS:2021yhb} (green dotted lines), where the observed limits agree with the median expected limits within 15\% and are within the regions containing 68\% of the distributions of the limits expected under the background-only hypothesis.}
\label{fig:limits_comp}
\end{figure*}

This search provides the first exclusions of hadronically decaying displaced tau leptons arising from LLPs with decay lengths smaller than $\approx$1\unit{m}.
The \sbb and \sdd observed limits are compared to other results~\cite{CMS:2020iwv,CMS:2021yhb} for representative signal points in Fig.~\ref{fig:limits_comp}.
Although the integrated luminosity of the data analyzed in this search is only $\approx$25\% of that used in the other searches, the obtained limits are much stronger, thanks to the new trigger, 
reconstruction, and machine-learning techniques.
The \sbb (\sdd) limits are better than those obtained previously by a factor of up to 10 (8). 

\begin{figure}[hbtp!]
\centering
\includegraphics[width=0.49\textwidth]{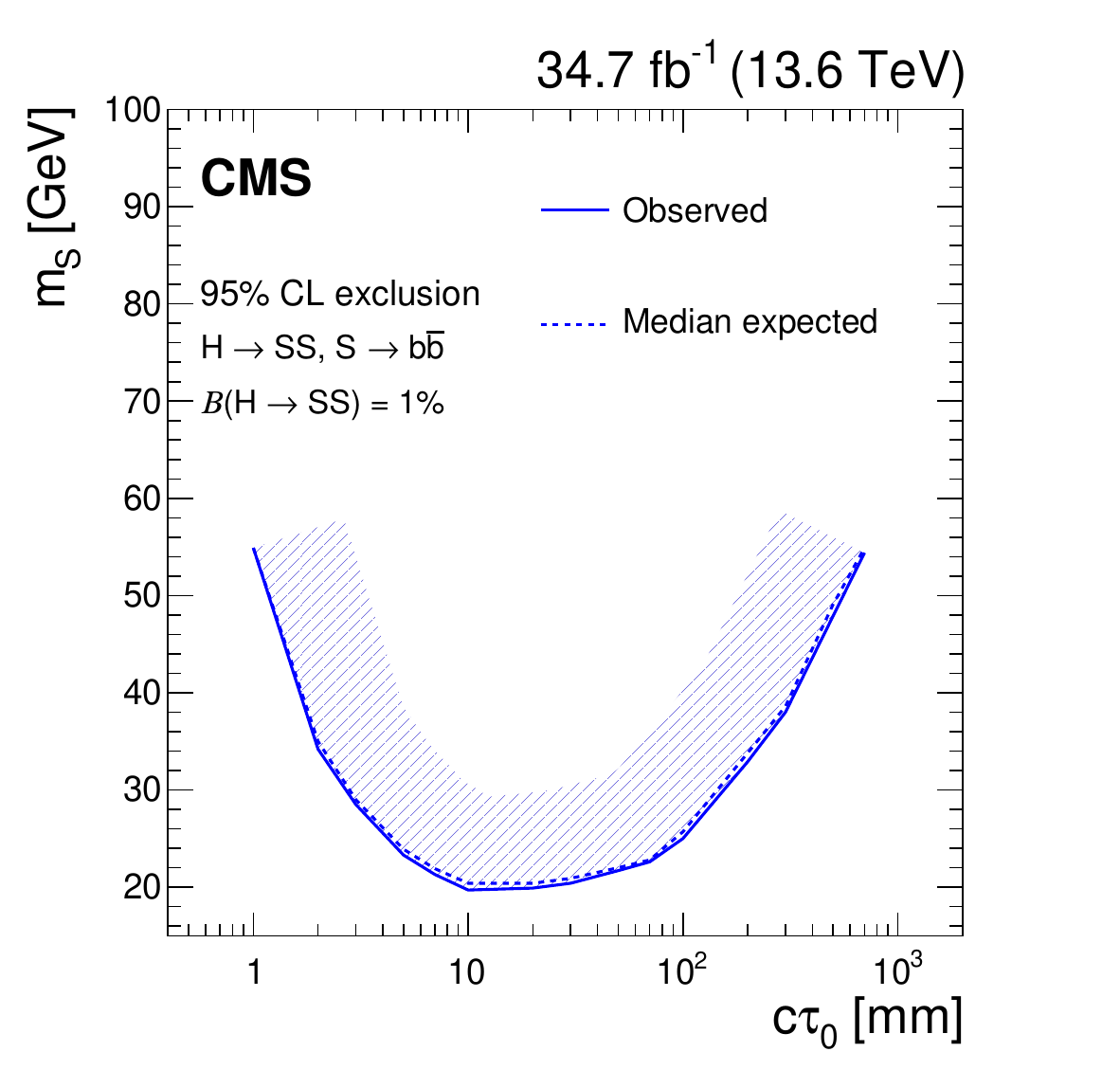}
\includegraphics[width=0.49\textwidth]{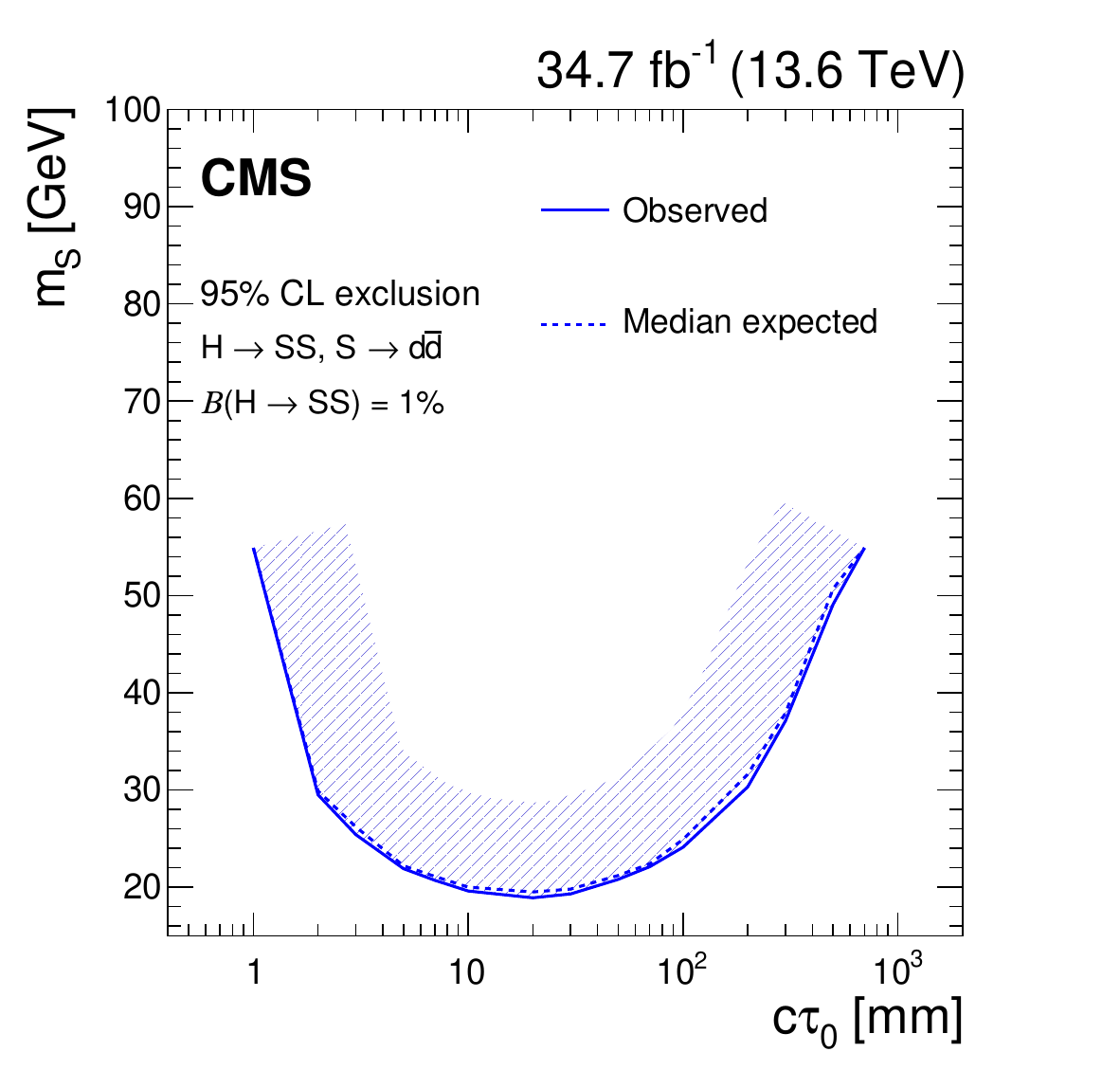}
\caption{The 95\% \CL limits on the LLP mass \mS for different proper decay lengths \ctau assuming a branching fraction of 1\% for the \HSS decay, and with subsequent \sbb (\cmsLeft) or \sdd (\cmsRight) decays. The solid (dashed) lines represent the observed 
(median expected) limits. The hashed areas indicate the direction of the excluded 
area from the observed limits.}
\label{fig:BR_exc}
\end{figure}

Figure~\ref{fig:BR_exc} shows the 95\% \CL limits on \mS for different \ctau assuming 
a branching fraction of 1\% for the \HSS decay, with subsequent \sbb or \sdd decays.
When \mS is larger than 40\GeV, \BHSS larger than 1\% is excluded for \ctau between 1.5 and 370\mm with \sbb decays, or for \ctau between 1.3 and 380\mm with \sdd decays.

\begin{figure}[hptb]
\includegraphics[width=0.49\textwidth]{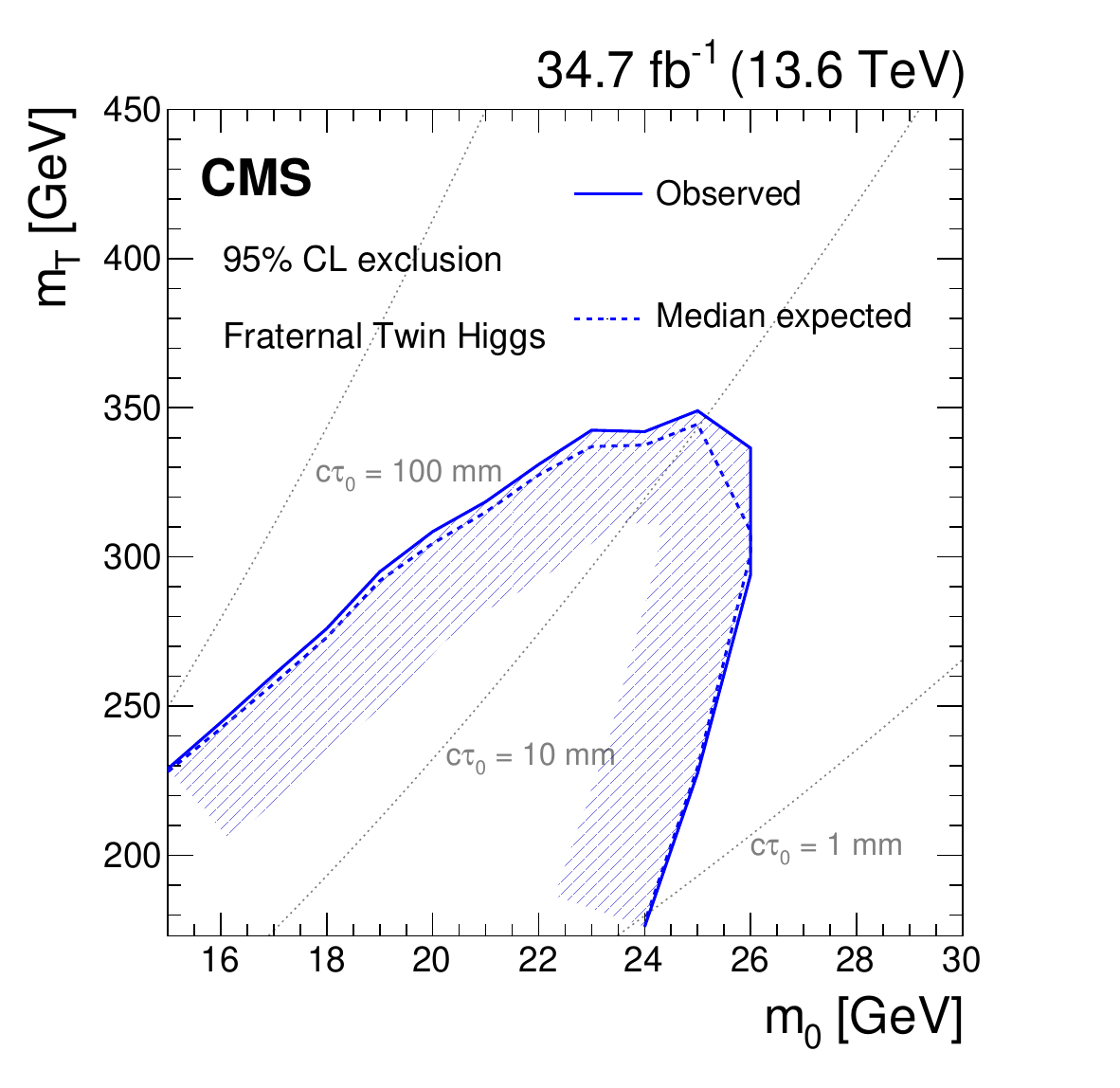}
\includegraphics[width=0.49\textwidth]{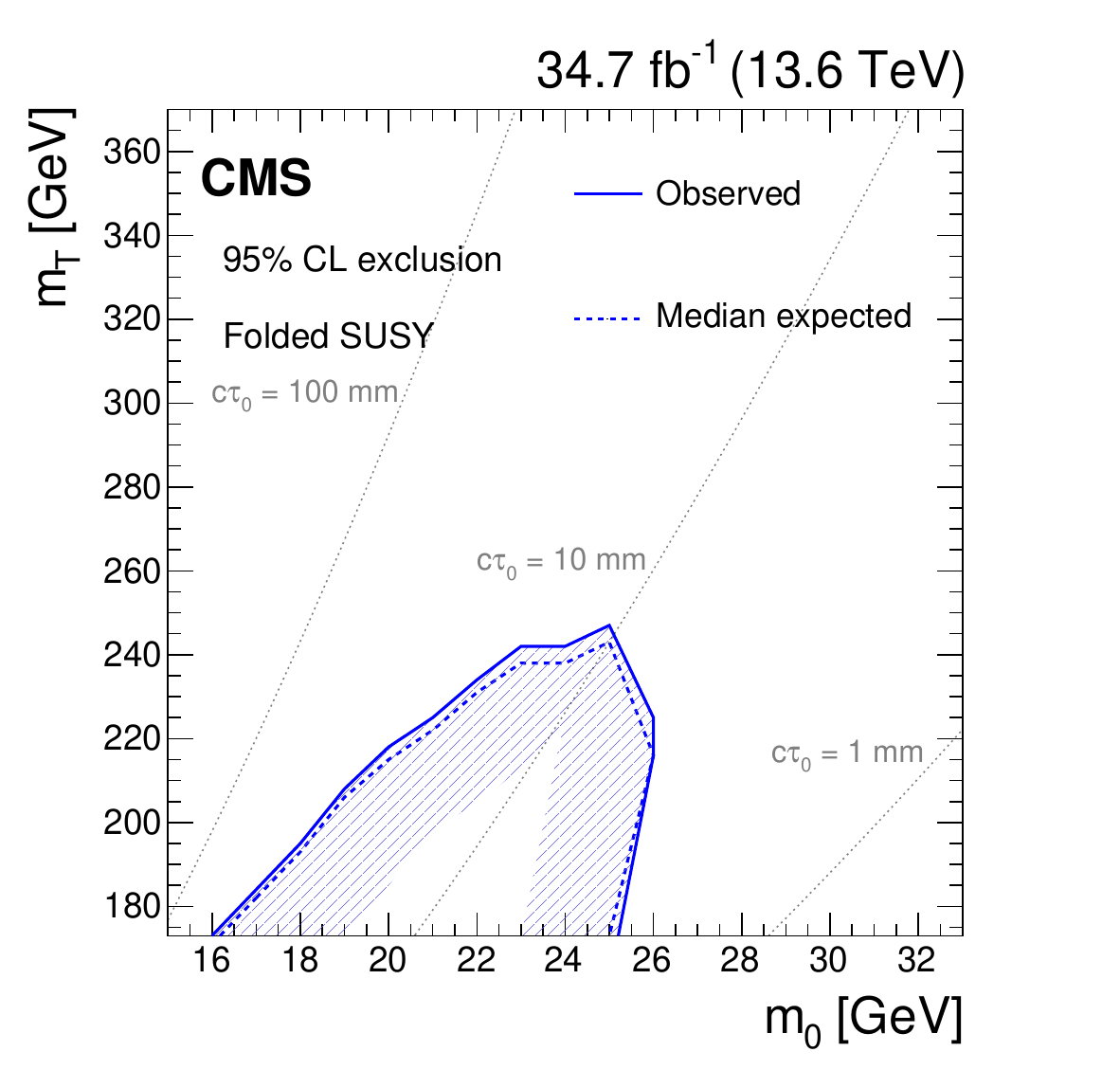}
\caption{The 95\% \CL limits on the dark-sector top quark partner mass \mTP for 
different hidden glueball masses \mG, in the fraternal twin Higgs model~\cite{Craig:2015pha} (\cmsLeft) 
and the folded SUSY model~\cite{Burdman:2006tz} (\cmsRight). The solid (dashed) lines represent the observed (median expected) limits. The hashed areas indicate the direction of the excluded area from the observed limits.}
\label{fig:NN_exc}
\end{figure}

{\tolerance=800 We also interpret the search with the fraternal twin Higgs (FTH)~\cite{Craig:2015pha} 
and folded SUSY (FSUSY)~\cite{Burdman:2006tz} models, which are two benchmarks for the neutral-naturalness scenario. For this interpretation, the scalar \cPS is interpreted as the 
lightest glueball \PG in the DS. The branching fraction for the 
Higgs boson decay to \PG (\BHGG) and the \ctau of \PG both depend on the glueball mass \mG and the mass \mTP of the top quark partner \PQT in the DS. 
These dependencies are taken from Ref.~\cite{deFlorian:2016spz}, assuming \BHGG is the same 
as the branching fraction for the Higgs boson to decay to hidden gluons multiplied by a phase space factor $\sqrt{\smash[b]{1-4\mG^{2}/m_{\PH}^{2}}}$.
For simplicity we also assume the branching fraction for $\PG\to\bbbar$ is 100\%,
since this is the dominant decay channel of \PG in the considered mass range because of the Higgs-portal interaction.
In this way, the \sbb limits are translated into 95\% 
\CL exclusions in the \mG-\mTP plane, as shown in Fig.~\ref{fig:NN_exc}. 
Top quark partner masses up to 350 (250)\GeV are excluded for the FTH (FSUSY) model.\par}

The data analyzed in this search was collected in the first year of the ongoing LHC Run~3.
The results in this paper already achieve an order-of-magnitude improvement over existing results, which represents a significant step forward in probing the phase space of exotic LLPs.
The full LHC Run-3 data set will correspond to a much larger integrated luminosity, which is expected to increase the sensitivity of this search significantly. The techniques introduced in this paper, together with newer techniques such as additional DJTs in a separate parking data stream dedicated to LLP searches~\cite{CMS:2024zhe}, will help realize the full potential of the complete Run-3 data set. Their future application to more challenging exotic LLP signatures will significantly enhance the discovery potential for BSM physics. 

\section{Summary}\label{sec:sum}

A search for light long-lived particles decaying into jets has been performed using proton-proton collision data 
corresponding to an integrated luminosity of 34.7\fbinv, collected with the CMS experiment at a center-of-mass energy of 13.6\TeV in 2022.
Novel techniques in trigger, reconstruction, and machine learning were developed for and employed in this search, leading to 
significant improvements over existing results.

The observed yields are consistent with the background predictions. The best limits to date are set 
for long-lived particles with masses between 15 and 55\GeV and with proper decay lengths smaller than $\approx$1\unit{m}. 
The search provides the first exclusions of hadronically decaying displaced tau leptons arising from LLPs with decay lengths smaller 
than $\approx$1\unit{m}. For the signature where the Higgs boson decays to two long-lived particles that further decay 
to bottom (down) quark pairs, branching fractions greater than 1\% for the exotic Higgs boson decay are excluded for a long-lived particle mass larger than 40\GeV and mean proper decay lengths between 1.5~(1.3) and 370~(380) \mm. For these signatures, the branching 
fraction limits are better than those obtained previously by a factor of up to 10~(8). Exclusions are also placed on the parameter space of the fraternal twin Higgs and folded supersymmetry models in the 
neutral naturalness scenario, giving lower limits on top quark partner masses of up to 350 and 250\GeV, respectively. The results are the first constraints placed on these models.

\begin{acknowledgments}
    We congratulate our colleagues in the CERN accelerator departments for the excellent performance of the LHC and thank the technical and administrative staffs at CERN and at other CMS institutes for their contributions to the success of the CMS effort. In addition, we gratefully acknowledge the computing centers and personnel of the Worldwide LHC Computing Grid and other centers for delivering so effectively the computing infrastructure essential to our analyses. Finally, we acknowledge the enduring support for the construction and operation of the LHC, the CMS detector, and the supporting computing infrastructure provided by the following funding agencies: SC (Armenia), BMBWF and FWF (Austria); FNRS and FWO (Belgium); CNPq, CAPES, FAPERJ, FAPERGS, and FAPESP (Brazil); MES and BNSF (Bulgaria); CERN; CAS, MoST, and NSFC (China); MINCIENCIAS (Colombia); MSES and CSF (Croatia); RIF (Cyprus); SENESCYT (Ecuador); ERC PRG, RVTT3 and MoER TK202 (Estonia); Academy of Finland, MEC, and HIP (Finland); CEA and CNRS/IN2P3 (France); SRNSF (Georgia); BMBF, DFG, and HGF (Germany); GSRI (Greece); NKFIH (Hungary); DAE and DST (India); IPM (Iran); SFI (Ireland); INFN (Italy); MSIP and NRF (Republic of Korea); MES (Latvia); LMTLT (Lithuania); MOE and UM (Malaysia); BUAP, CINVESTAV, CONACYT, LNS, SEP, and UASLP-FAI (Mexico); MOS (Montenegro); MBIE (New Zealand); PAEC (Pakistan); MES and NSC (Poland); FCT (Portugal);  MESTD (Serbia); MCIN/AEI and PCTI (Spain); MOSTR (Sri Lanka); Swiss Funding Agencies (Switzerland); MST (Taipei); MHESI and NSTDA (Thailand); TUBITAK and TENMAK (Turkey); NASU (Ukraine); STFC (United Kingdom); DOE and NSF (USA).
    \hyphenation{Rachada-pisek} Individuals have received support from the Marie-Curie program and the European Research Council and Horizon 2020 Grant, contract Nos.\ 675440, 724704, 752730, 758316, 765710, 824093, 101115353, 101002207, and COST Action CA16108 (European Union); the Leventis Foundation; the Alfred P.\ Sloan Foundation; the Alexander von Humboldt Foundation; the Science Committee, project no. 22rl-037 (Armenia); the Belgian Federal Science Policy Office; the Fonds pour la Formation \`a la Recherche dans l'Industrie et dans l'Agriculture (FRIA-Belgium); the F.R.S.-FNRS and FWO (Belgium) under the ``Excellence of Science -- EOS" -- be.h project n.\ 30820817; the Beijing Municipal Science \& Technology Commission, No. Z191100007219010 and Fundamental Research Funds for the Central Universities (China); the Ministry of Education, Youth and Sports (MEYS) of the Czech Republic; the Shota Rustaveli National Science Foundation, grant FR-22-985 (Georgia); the Deutsche Forschungsgemeinschaft (DFG), among others, under Germany's Excellence Strategy -- EXC 2121 ``Quantum Universe" -- 390833306, and under project number 400140256 - GRK2497; the Hellenic Foundation for Research and Innovation (HFRI), Project Number 2288 (Greece); the Hungarian Academy of Sciences, the New National Excellence Program - \'UNKP, the NKFIH research grants K 131991, K 133046, K 138136, K 143460, K 143477, K 146913, K 146914, K 147048, 2020-2.2.1-ED-2021-00181, and TKP2021-NKTA-64 (Hungary); the Council of Science and Industrial Research, India; ICSC -- National Research Center for High Performance Computing, Big Data and Quantum Computing and FAIR -- Future Artificial Intelligence Research, funded by the NextGenerationEU program (Italy); the Latvian Council of Science; the Ministry of Education and Science, project no. 2022/WK/14, and the National Science Center, contracts Opus 2021/41/B/ST2/01369 and 2021/43/B/ST2/01552 (Poland); the Funda\c{c}\~ao para a Ci\^encia e a Tecnologia, grant CEECIND/01334/2018 (Portugal); the National Priorities Research Program by Qatar National Research Fund; MCIN/AEI/10.13039/501100011033, ERDF ``a way of making Europe", and the Programa Estatal de Fomento de la Investigaci{\'o}n Cient{\'i}fica y T{\'e}cnica de Excelencia Mar\'{\i}a de Maeztu, grant MDM-2017-0765 and Programa Severo Ochoa del Principado de Asturias (Spain); the Chulalongkorn Academic into Its 2nd Century Project Advancement Project, and the National Science, Research and Innovation Fund via the Program Management Unit for Human Resources \& Institutional Development, Research and Innovation, grant B39G670016 (Thailand); the Kavli Foundation; the Nvidia Corporation; the SuperMicro Corporation; the Welch Foundation, contract C-1845; and the Weston Havens Foundation (USA).
\end{acknowledgments}

\bibliography{auto_generated}

\clearpage
\appendix

\section{Glossary of acronyms}
\label{sec:glossary}

\begin{tabular}{ll}
AVF & Adaptive vertex fitter\\
AVR & Adaptive vertex reconstruction\\
BSM & Beyond the standard model\\
\CL & Confidence level\\
DJ & Displaced jet\\
DJT & Displaced-jet trigger\\
DS & Dark sector\\
DV & Displaced vertex\\
ECAL & Electromagnetic calorimeter\\
FTH & Fraternal twin Higgs\\
FSUSY & Folded supersymmetry\\
GNN & Graph neural network\\
HCAL & Hadron calorimeter\\
HLT & High-level trigger\\
LLP & Long-lived particle \\
LO & Leading order\\
LSTM & Long short-term memory\\
PCA & Point of the closest approach\\
PV & Primary vertex\\
QCD & Quantum chromodynamics\\
SM & Standard model\\
SR & Signal region\\
SUSY & Supersymmetry\\

\end{tabular}

\cleardoublepage \section{The CMS Collaboration \label{app:collab}}\begin{sloppypar}\hyphenpenalty=5000\widowpenalty=500\clubpenalty=5000
\cmsinstitute{Yerevan Physics Institute, Yerevan, Armenia}
{\tolerance=6000
A.~Hayrapetyan, A.~Tumasyan\cmsAuthorMark{1}\cmsorcid{0009-0000-0684-6742}
\par}
\cmsinstitute{Institut f\"{u}r Hochenergiephysik, Vienna, Austria}
{\tolerance=6000
W.~Adam\cmsorcid{0000-0001-9099-4341}, J.W.~Andrejkovic, T.~Bergauer\cmsorcid{0000-0002-5786-0293}, S.~Chatterjee\cmsorcid{0000-0003-2660-0349}, K.~Damanakis\cmsorcid{0000-0001-5389-2872}, M.~Dragicevic\cmsorcid{0000-0003-1967-6783}, P.S.~Hussain\cmsorcid{0000-0002-4825-5278}, M.~Jeitler\cmsAuthorMark{2}\cmsorcid{0000-0002-5141-9560}, N.~Krammer\cmsorcid{0000-0002-0548-0985}, A.~Li\cmsorcid{0000-0002-4547-116X}, D.~Liko\cmsorcid{0000-0002-3380-473X}, I.~Mikulec\cmsorcid{0000-0003-0385-2746}, J.~Schieck\cmsAuthorMark{2}\cmsorcid{0000-0002-1058-8093}, R.~Sch\"{o}fbeck\cmsorcid{0000-0002-2332-8784}, D.~Schwarz\cmsorcid{0000-0002-3821-7331}, M.~Sonawane\cmsorcid{0000-0003-0510-7010}, W.~Waltenberger\cmsorcid{0000-0002-6215-7228}, C.-E.~Wulz\cmsAuthorMark{2}\cmsorcid{0000-0001-9226-5812}
% ...
\par}
\vskip\cmsinstskip
\dag:~Deceased\\
$^{1}$Also at Yerevan State University, Yerevan, Armenia\\
$^{2}$Also at TU Wien, Vienna, Austria\\
% ...
\end{sloppypar}
\end{document}